\documentclass[a4paper]{jpconf}
\usepackage{amssymb,hyperref,units,graphicx,txfonts,floatflt}
\newcommand{\un}[1]{\,\mathrm{#1}}

\newcommand{\degC}[1]{\mbox{$#1\,^\circ{\mathrm C}$}}
\newcommand{\inch}{\texttt{"}} 
\begin{document}
\title{Supernova Neutrino Detection with IceCube}
\author{Lutz K\"opke for the IceCube Collaboration}
\address{University of Mainz, D-55099 Mainz, Germany}
\ead{lutz.koepke@uni-mainz.de}
\begin{abstract}
IceCube was completed in December 2010. It 
forms a lattice of 5160 photomultiplier tubes that monitor a volume of $\sim$\,\unit[1]{km$^3$} in the deep Antarctic ice for particle induced photons. The telescope was designed to detect neutrinos with energies greater than 100 GeV. Owing to subfreezing ice temperatures, 
the photomultiplier dark noise rates are particularly low. Hence IceCube can also detect large numbers of MeV neutrinos by observing a collective rise in all photomultiplier rates on top of the dark noise. 
With 2 ms timing resolution, IceCube can track subtle features in the 
temporal development of the supernova neutrino burst. For a supernova at the galactic center, its sensitivity matches that of a background-free megaton-scale supernova search experiment. The sensitivity decreases to 20 standard deviations at the galactic edge (30 kpc) and 6 standard deviations at the Large Magellanic Cloud (50 kpc). IceCube is sending triggers from potential supernovae to the Supernova Early Warning System. 
The sensitivity to neutrino properties such as the neutrino hierarchy is discussed and simulations of
tantalizing signatures, such as the formation of a quark star or a black hole as well as the characteristics of shock waves are presented.
All results are preliminary.
\end{abstract}

%
%
\section{Introduction}\label{sec:intro}
It was recognized early by~\cite{Pryor} and~\cite{PhysRevD.53.7359} that neutrino telescopes offer the possibility to monitor our Galaxy for supernovae. IceCube is uniquely suited for this measurement due to its location and 1 km$^3$ size. The noise rates in IceCube's photomultiplier tubes average around 540 Hz since they are surrounded by inert and cold ice with depth dependent temperatures ranging from \degC{-43} to \degC{-20}. At depths between \unit[(1450 -- 2450)]{m} they are partly shielded from cosmic rays.
Cherenkov light induced by neutrino interactions will increase the count rate of all light sensors above their average value. 
Although this increase in individual light sensor is not statistically significant, the effect will be clearly seen once the rise is considered collectively over many sensors.
 The 5160 photomultipliers are sufficiently far apart such that the probability to detect light from a single interaction in more than 
one DOM is small. With absorption lengths exceeding 100 m, photons travel long distances in the ice such that each DOM effectively monitors several hundred cubic-meters of ice. The inverse beta process  $\bar \nu_\mathrm{e} + \mathrm{p} \rightarrow \mathrm{e^+} + \mathrm{n}  $  dominates supernova neutrino interactions with $ \mathcal{O}(10\un{MeV})$ energy in ice or water, leading to positron tracks of about $0.6 \un{cm} \cdot E_\nu/ \un{MeV}$ length.
Considering the approximate $E_\nu^2$ dependence of the cross section, the light yield per neutrino roughly scales with $E_\nu^3$. The detection principle was demonstrated with the AMANDA experiment, IceCube's predecessor~\cite{bib:AMANDAold}.
Since 2009, IceCube has been sending real-time datagrams to the Supernova Early Warning System (SNEWS)~\cite{antonioli-2004-6} when detecting supernova candidate events.

\section{Detector}\label{sec:detector}
The Digital Optical Module (DOM)
is the fundamental element in the IceCube architecture. Housed in a 13\inch{} (\unit[33]{cm}) borosilicate glass pressure sphere, it contains a Hamamatsu 10\inch{} hemispherical photomultiplier tube~\cite{ic3:pmt-paper} as well as several electronics boards containing a processor, memory, flash file system and realtime operating system that allows each DOM to operate as a complete and autonomous data acquisition system~\cite{daq}. It stores the digitized data internally and transmits the information to a surface data acquisition system on request. The supernova detection relies on continuous measurements of photomultiplier rates. The rate information is stored and buffered on each DOM in a 4-bit counter in \unit[1.6384]{ms} time bins ($2^{16}$ cycles of the 40 MHz clock). For the real-time processing, the information is synchronized with the help of a GPS clock and regrouped in \unit[2]{ms} bins. The South Pole is out of reach for most communication satellites and high bandwidth connectivity is available only for about 6 hours per day. 
Therefore, a dedicated Iridium-satellite~\cite{IRIDIUM} connection is used by the SNDAQ host system to transmit urgent alerts. In that case, a short datagram is sent to the northern hemisphere. The receiving system parses the message and forwards information on the supernova candidate event to the international SNEWS group. The time delay between photons hitting the optical module and the arrival of the datagram at SNEWS stands at about \unit[6]{min}, providing close to real-time monitoring and triggering. 
Due to satellite bandwidth constraints, the data are re-binned in \unit[0.5]{s} intervals and then subjected to a statistical
 online analysis; the fine time information in \unit[2]{ms} intervals is transmitted for a period starting 30 s before and ending 60 s after a trigger flagging a candidate supernova explosion. 

\begin{figure}[Ht]
\centering
\includegraphics[angle=0,width=0.485\textwidth]{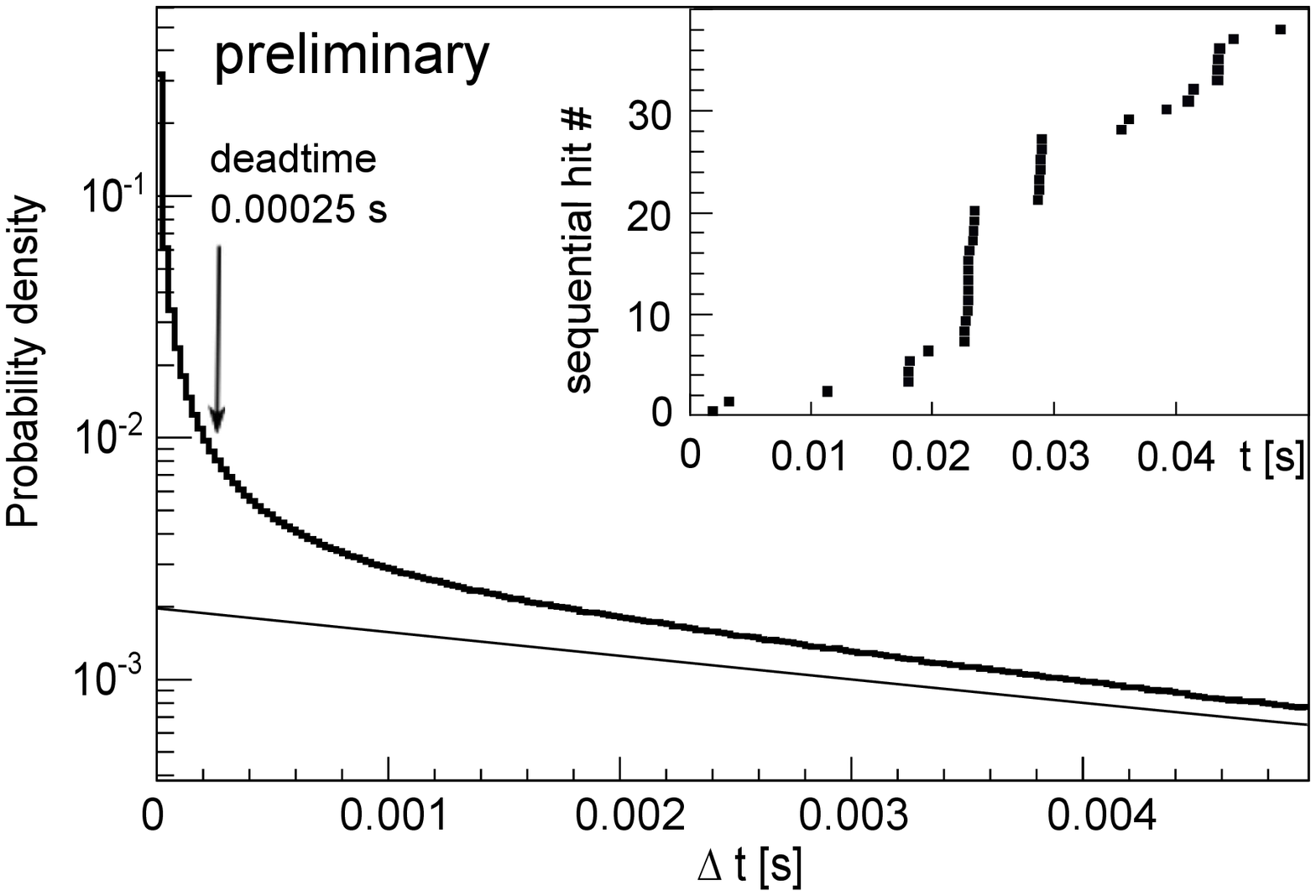}\quad\quad
\includegraphics[angle=0,width=0.445\textwidth]{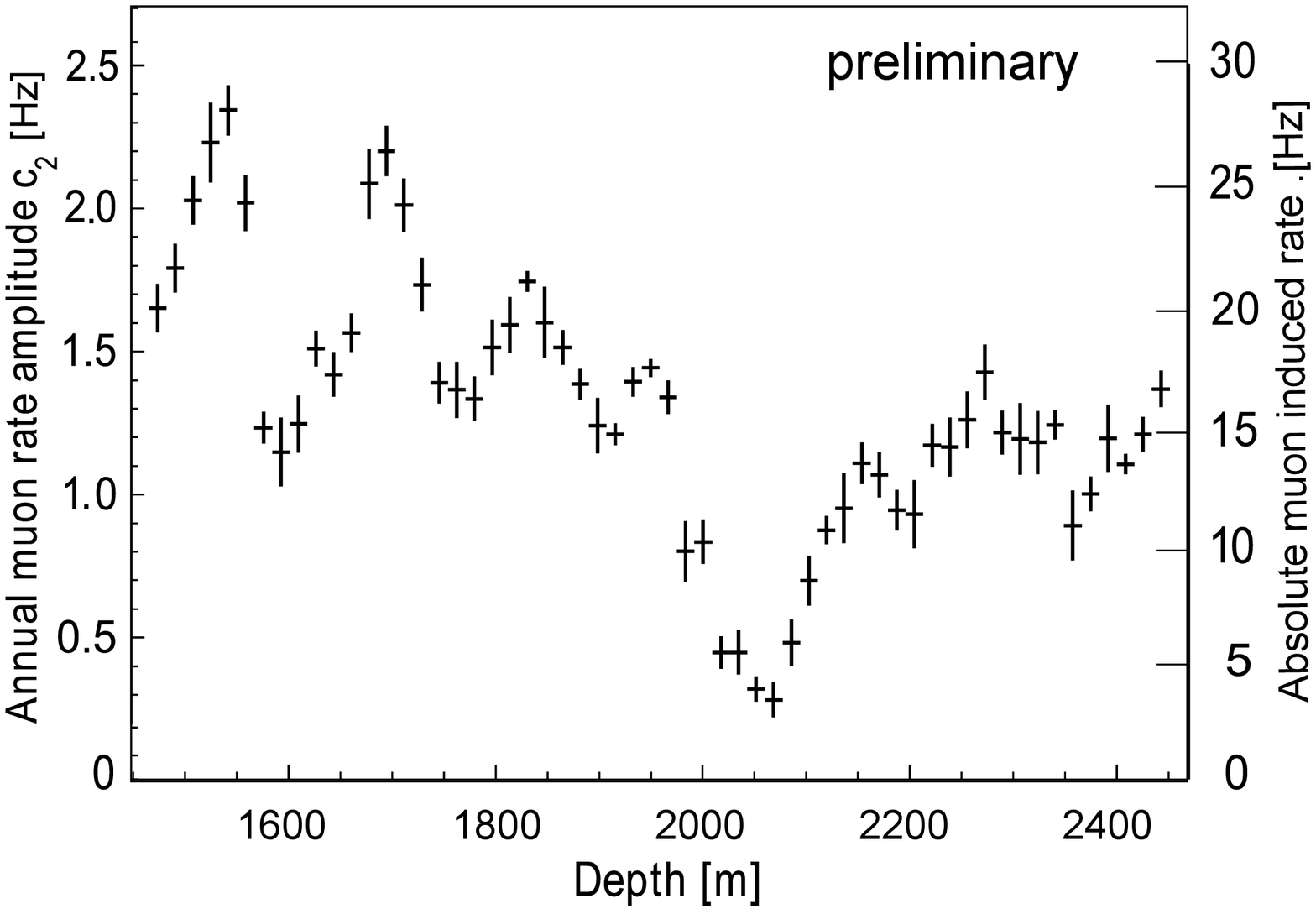}
\caption{Left: Probability density distribution of time differences between pulses for noise (bold line) and the expectation for a Poissonian process fitted in the range 15 ms $ < \Delta T < $ 50 ms (thin line). The excess is due to bursts of correlated hits, as indicated by the 50 ms long snapshot of hit times shown in the insert. Right: Parameter $c_2$ (see section~\ref{sec:detector_performance}) and estimated muon induced rate as function of depth.  The variation with depth is mostly due to the optical properties of the ice and muons ranging out.}
\label{fig:dom-noise-response1}
\end{figure}

Several effects contribute to the prevailing low noise rate of 540 Hz: a Poissonian noise contribution from radioactivity, atmospheric muons and remaining thermal noise, 
as well as correlated noise from Cherenkov radiation and scintillation originating in the glass of the photomultiplier and the pressure vessel. 
 It is suspected that residual $\beta$ and $\alpha$ decays from trace elements in the uranium/thorium chain powers cerium-based scintillation, causing a series of pulses.
The observed time difference between noise hits deviates from an exponential distribution as expected for a Poissonian process (see Fig.~\ref{fig:dom-noise-response1}). With typical times between correlated noise pulses of $\mathcal{O}(\unit[100]{\mu s})$, the signal-to-noise ratio of the measurement can be improved by adding an artificial dead time that is configurable by a field programmable gate array in the DOM. The optimal setting for the dead time with respect to the signal over noise ratio for supernovae was found to be $\tau \approx \unit[250]{\mu s}$.

\section{Real-time analysis method}
\label{sec:data_analysis}

Counting $N_i$ pulses during a given time interval $\Delta t$, rates $r_i=N_i/\Delta t$ 
for DOM $i$, are derived.  The index $i$ ranges from $1$ to the total number of operational optical modules $N_\mathrm{DOM}$. With sufficiently large $\Delta t$'s, the distributions of the $r_i$'s can be described by lognormal distributions  (see Fig.~\ref{fig:rates}) that, for simplicity, are approximated by Gaussian distributions with rate expectation values 
$\langle r_i\rangle$ and corresponding standard deviation expectation values $\langle\sigma_i\rangle$.
These 
expectation values are computed from moving \unit[300]{s} time intervals before and after the investigated time interval. At the beginning and the end of a SNDAQ-run, asymmetric intervals are used.  The time windows exclude $30\un{s}$ before and after the investigated bin in order to reduce the impact of a wide signal on the mean rates. 
 The most likely \textit{collective rate deviation} $\Delta\mu$ of all DOM noise rates $r_i$ from their individual $\langle r_i\rangle$'s, assuming the null hypothesis of no signal, is obtained by maximizing the likelihood
\begin{equation}
	\mathcal{L}(\Delta\mu) = \prod_{i=1}^{N_\mathrm{DOM}} \, \frac{1}{\sqrt{2\pi}\,\langle\sigma_i\rangle} \, {\rm exp}(-\frac{(r_i-(\langle r_i\rangle+\epsilon_i\,\Delta\mu))^2}{2\langle\sigma_i\rangle^2}) \,\, .
\end{equation}
Here $\epsilon_i$ denotes a correction for module and depth dependent detection probabilities. An analytic minimization of $-\ln\mathcal{L}$ leads to
\begin{equation}
\Delta\mu = \sigma_{\Delta\mu}^2 \sum_{i=1}^{N_\mathrm{DOM}} \, \frac{\epsilon_i\,(r_i - \langle r_i\rangle)}{\langle\sigma_i\rangle^2} \quad \mathrm{with}\quad \sigma_{\Delta\mu}^2 = \left(\sum_{i=1}^{N_\mathrm{DOM}} \, \frac{{\epsilon_i}^2}{\langle\sigma_i\rangle^2}\right)^{-1} \quad .
\end{equation}
Assuming uncorrelated background noise and a large number of contributing DOMs, the significance $\xi = \Delta\mu/\sigma_{\Delta\mu}$ should approximately follow a Gaussian distribution with unit width centered at zero.  The likelihood that a deviation is caused by an isotropic and homogeneous illumination of the ice can be calculated from the $\chi^2$-probability
$-2\ln({\mathcal L})=\chi_{\Delta\mu}^2 = \sum_{i=1}^{N_\mathrm{DOM}} ({r_i-(\langle r_i\rangle+\epsilon_i\,\Delta\mu)/\langle\sigma_i\rangle})^2$.

To cover model uncertainties, analyses with time bases of \unit[0.5]{s}, \unit[4]{s} and \unit[10]{s} are run in parallel.
The collective rate deviation $\Delta\mu$ and its uncertainty $\sigma_{\Delta\mu}$ in the time bases of \unit[4]{s} and \unit[10]{s} are calculated using sliding windows in \unit[0.5]{s} steps and extracting the maximal significance. This procedure ensures that the signal detection efficiency is not reduced by binning effects. 
\section{Detector performance}\label{sec:detector_performance}
\begin{figure}
\centering
\includegraphics[angle=0,width=0.46\textwidth]{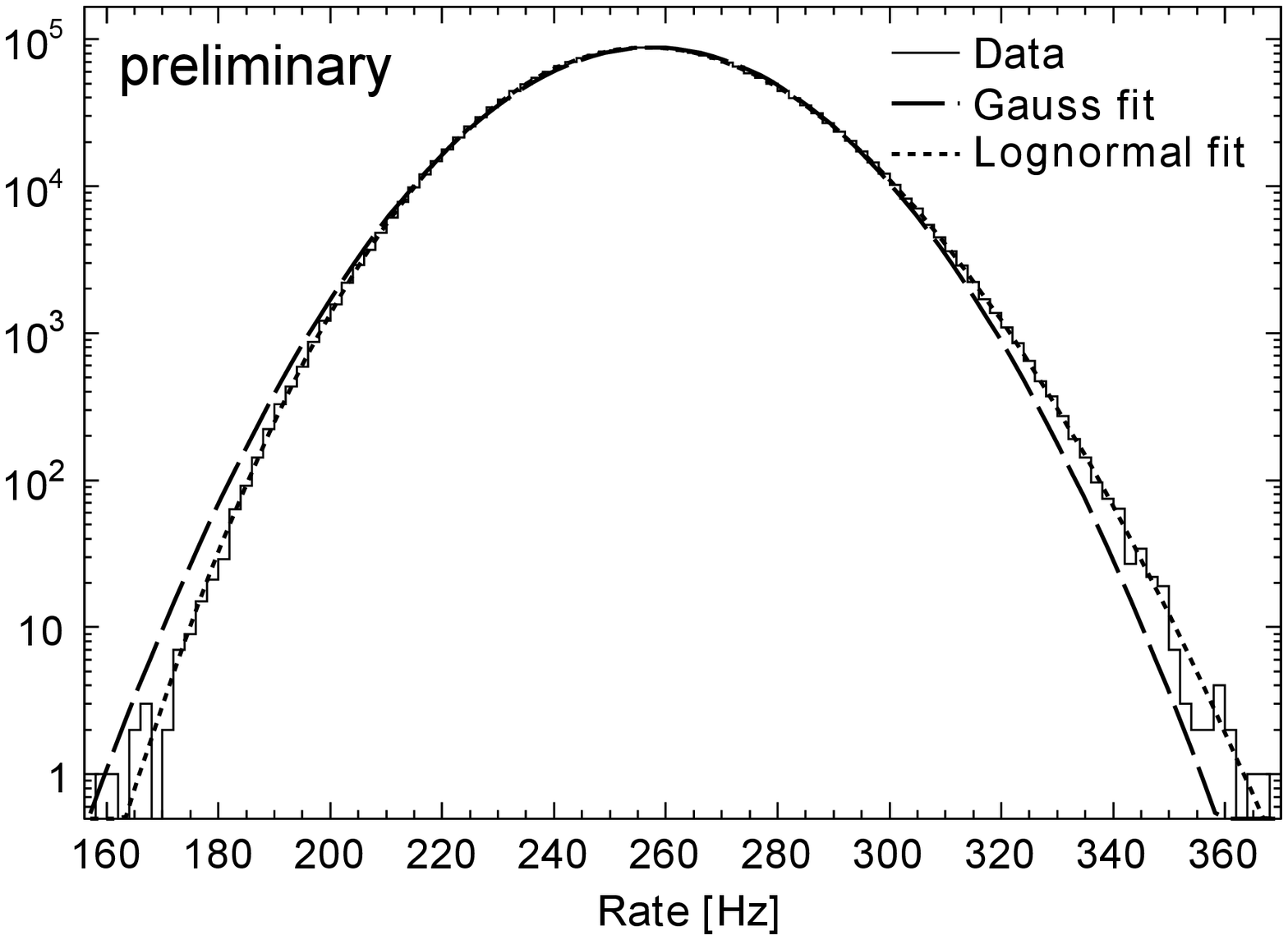} \quad\quad
\includegraphics[angle=0,width=0.45\textwidth]{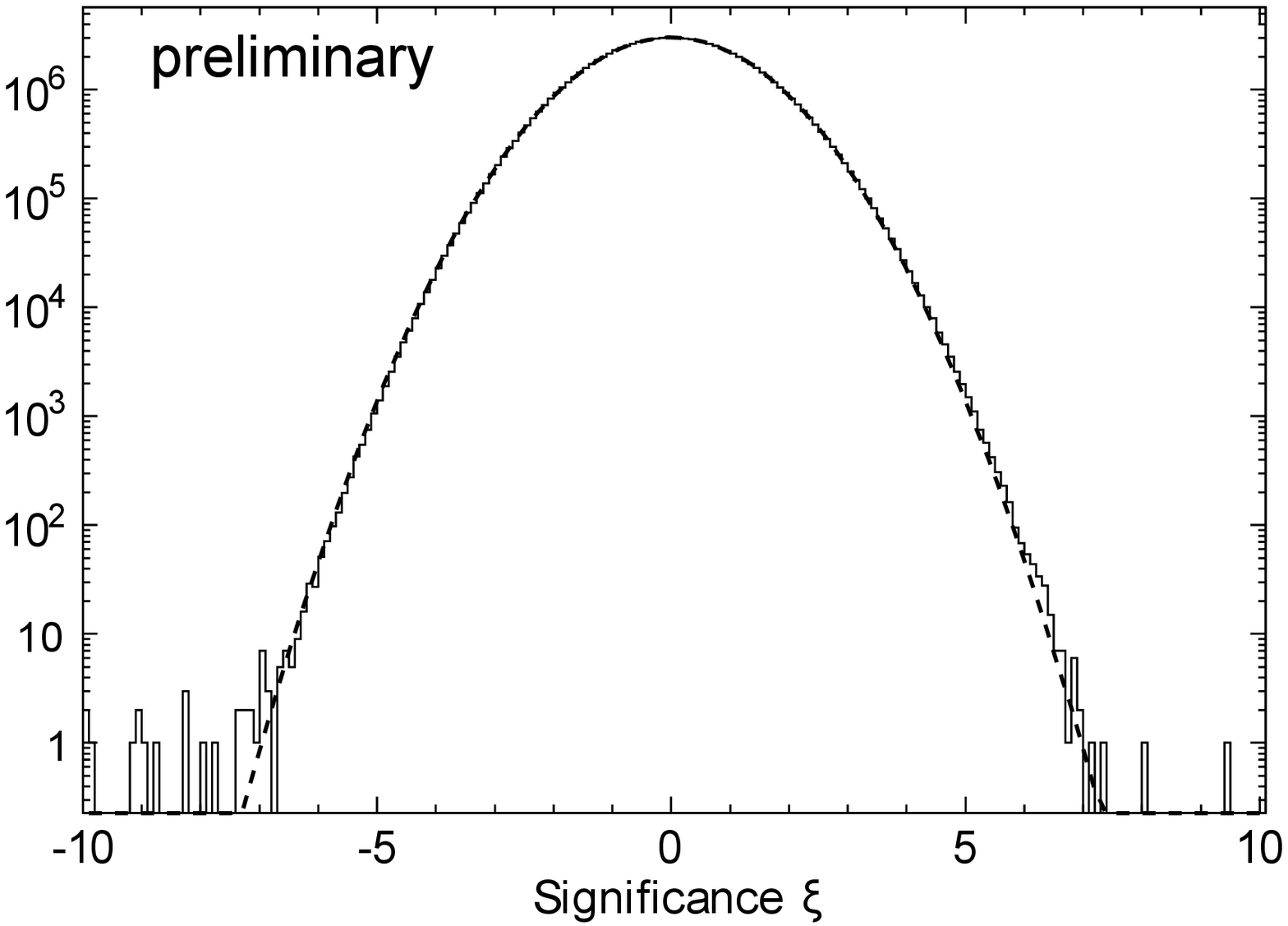}
\caption{Left: Rate distribution of a typical standard efficiency DOM taken over 29 consecutive days.  Each measurement corresponds to 0.5 s integration time. The application of an  artificial deadtime of 250 $\mu$s lowers the mean rate of this DOM to $\approx$ 260 Hz. Gaussian and lognormal fits are shown. Right: Significance distribution in \unit[0.5]{s} binning for a detector uptime of 556 days with 22 and 40 strings deployed. The two outliers at $\xi$= 8 and 9.5 occurred during test runs employing artificial light. The dashed line shows 
a Gaussian fit with $\sigma = 1.27$.%
}\label{fig:rates}
\end{figure}

The DOM rates $r(t)$ are characterized by an exponential rate decrease over long time periods and a slight seasonal modulation that is represented quite well by the formula $r(t) = r_0 + c_1 e^{-t/\tau} + c_2 \sin(2\pi (t/\mathrm{year}))$. Interpreting the seasonal modulation as being due to stratospheric temperature variations, the averaged muonic contribution to single DOM rates is $\approx$ \unit[16]{Hz}, with a strong depth dependence due to variations in absorption length and muons ranging out (see right plot of Fig.~\ref{fig:dom-noise-response1}). 
The slightly skewed rate distribution of a single DOM is better described by a lognormal distribution than by a Gaussian (see left plot of Fig.~\ref{fig:rates} with \unit[250]{$\mu$s} dead time applied).  Thanks to the tight quality control, the average noise rates between DOMs vary only by 10\%.  

The expected signal sensitivity in IceCube is somewhat reduced due to two types of correlations between pulses that introduce supra-Poissonian fluctuations. The first correlation concerns single photomultiplier tubes and arises from a burst of photons produced by radioactive decays in the pressure vessel discussed above. The artificial deadtime suppresses - but does not completely remove - the effect. This leads to a broadening of a factor of $\approx 1.28$ in the standard deviation of single photomultiplier tubes w.r.t. to the Poissonian expectation. The second correlation arises from the cosmic-ray muon background discussed above: a single cosmic ray shower can produce a bundle of muons which is seen by hundreds of optical modules. As evident from the right plot of Fig.~\ref{fig:rates}, the measured significance distribution $\xi$ is broader than expected and can be fairly well fitted by a Gaussian with width $\sigma = 1.27$. The broadening increases with the size of the detector and has reached $\sigma = 1.43$ with 79 operating strings. Offline, roughly half of the hits associated with triggered cosmic ray muon can be removed, lowering the broadening to $\sigma = 1.06$. The standard deviation of the rate sum of all DOMs suffers from both effects and turns out to be between $1.3 -1.7$ times larger than the Poissonian expectation for 2 ms and 500 ms bins, respectively. Subtracting hits associated with cosmic ray muons lowers the standard deviation to $(1.24 - 1.32)\sqrt{\sum_i r_i}$, slightly dependent on the binning. It may be possible to further reduce this broadening in the future using a better identification of bursts of correlated hits. 
\section{Production, interaction and detection of supernova neutrinos}\label{sec:theory}
We use the Lawrence-Livermore~\cite{AstropPhys.496.216} and Garching models~\cite{AstrAstrop.450.1} as benchmarks, but will also refer to more specific models that were selected to demonstrate IceCube's physics performance. The spherically symmetric Lawrence-Livermore simulation was performed from the onset of the collapse to \unit[18]{s} after the core bounce, encompassing the complete accretion phase and a large part of the cooling phase. It is modeled after SN 1987A and assumes a 20 $M_\odot$ progenitor. The total emitted energy is $\unit[2.9\,\times\,10^{53}]{erg}$, of which 16\,\% is carried by $\bar\nu_\mathrm{e}$ with \unit[15.3]{MeV} energy on average. The newer spherically symmetric Garching simulations include more detailed information on neutrino energy spectra and use a sophisticated neutrino transport mechanism. They cover \unit[0.80]{s} following the collapse of an O-Ne-Mg \unit[8 -- 10]{} $M_\odot$ progenitor star, that is destabilized due to rapid electron capture on neon and magnesium. This class of stars may represent up to 30\,\% of all core collapse supernovae.

Neutrino oscillations in the supernova environment play an important role, modify the detected rate and thus provide sensitivity on neutrino properties. Neutrinos streaming out of the core will encounter matter densities ranging from \unit[$10^{13}$]{kg/m$^3$} to zero and through two MSW-resonance layers. Both mix the initial fluxes of $\nu_\mathrm{e}$, $\bar\nu_\mathrm{e}$ and all other flavors $\nu_\mathrm{x}$ depending on the survival probabilities. 
We consider two limiting cases as benchmarks to discuss the effect of the assumed neutrino hierarchy on the spectra observed with IceCube. Scenario A describes the normal neutrino hierarchy case and Scenario B represents the inverted hierarchy case with a static density profile of the supernova, both paired with a relatively large mixing angle $\theta_{13} > 0.9^\circ$. The oscillation scenario B for an inverted neutrino mass hierarchy shows the largest signal for the Lawrence-Livermore and Garching models because energetic $\bar\nu_\mathrm{x}$ will oscillate into $\bar\nu_\mathrm{e}$, harden their spectrum and thus increase the detection probability. The scenario without any oscillation is given as a reference and leads typically to the weakest signal.

In the simulation we include $\nu_\mathrm{e}$ and $\bar{\nu}_\mathrm{e}$ interactions on protons, electrons and $^{16/17/18}\mathrm{O}$ as well as positron annihilation and neutron capture. 
Reactions producing electrons or positrons in the final state radiate $N_{\gamma}=325.4$ cm$^{-1}\cdot x$ ($x$ in cm)
Cherenkov photons in the (300 - 600) nm wavelength range along their flight path $x$, as long as their kinetic energies exceed the Cherenkov threshold of 0.272 MeV. The mean travel path for $\mathcal{O}(\unit[10]{MeV})$ from $\bar\nu_\mathrm{e}$, including secondary leptons with energies above the Cherenkov threshold as well as  positron annihilation, was determined to be $\bar{x} = (0.577 \pm 0.005 \,\rm{(stat.)}\pm 0.029 \,\rm{(syst.)})\un{cm}$ $\cdot E_{\mathrm{e}^+}/\un{MeV}$. 

Neutrinos are detected in IceCube by registering predominantly single Cherenkov photons radiated by the neutrino interaction products. 
The optical scattering and absorption in glacial ice at the South Pole has been studied extensively~\cite{iceproperties06}. 
The mean number of photons recorded by an optical module averaged over energy is given by 
$N^\mathrm{detect}_\gamma = \epsilon_\mathrm{dead time}\cdot n^{\mathrm{interact}}_\nu\,\cdot\,\overline{N_{\gamma}}\cdot V^\mathrm{eff}_\mathrm{\gamma}$, 
where $n^{\rm interact}_\nu$ is the neutrino density, $\epsilon_\mathrm{dead time}\approx 0.87/(1+r_{SN}\cdot \tau)$ is the loss in signal rate $r_{SN}$ due to the artificial deadtime of $\tau=250$ $\mu$s, $\overline{N_{\gamma}}$ is the energy averaged number of Cherenkov photons, and 
$V^\mathrm{eff}_\mathrm{\gamma}$ is the so-called single photon effective volume, that varies strongly with the photon absorption. As a first approximation, it can be estimated by the product of average Cherenkov spectrum and DOM sensitivity weighted absorption length ($\approx \,$\unit[100]{m}), DOM geometric cross section (\unit[0.0856]{m$^2$}), Cherenkov spectrum weighted optical module sensitivity ($\approx \,$0.071), average angular sensitivity including cable shadowing effects ($\approx\,$0.32), and the fraction of single photon hits passing the electronic DOM threshold ($\approx \,$0.85). 
\begin{floatingfigure}[l]{8cm}
\includegraphics[width=7cm, bb=0 -1 501 508]{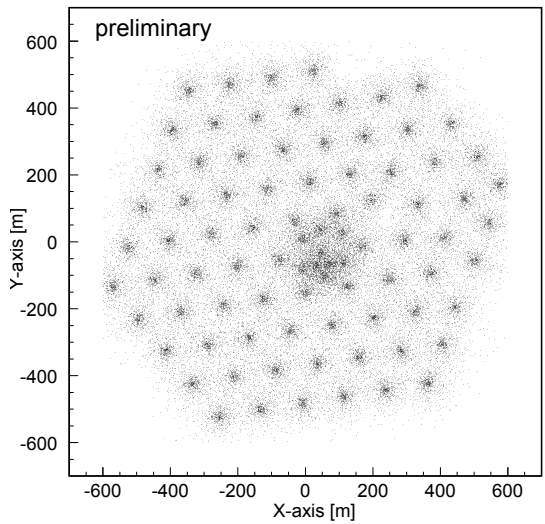}
\caption{\label{xyplot} Detected neutrino inverse beta decay interaction vertices projected onto horizontal plane (GEANT simulation with 10 million neutrino interactions). }
\end{floatingfigure}
We use two alternative procedures to calculate the number of detected signal hits from the number of neutrinos crossing the detector: the first approach relies on predetermined tables \cite{Lundberg:2007yg}, created to track photons across the Antarctic ice and separates simulations of particle interactions, Cherenkov photon creation, propagation and detection. We obtain $V_{\gamma}^\mathrm{eff} = 0.160 \pm 0.004 \,\rm{(stat.)} \pm 0.020 \,\rm{ (syst.)} \un{m^3}$. 
For positrons with a cross section weighted average energy of $\overline{E_\mathrm{e^+}}$=20 MeV  one obtains $\overline{N_{\gamma}}\cdot V_{\gamma}^\mathrm{eff} \approx (590\pm 80)\, \mathrm{m^3}$. This volume corresponds to an envisioned sphere of $\approx 5.2\un{m}$ radius centered at the optical module position, with full sensitivity inside and zero outside. To give an example, a study of the initial 380 ms of the burst in the Lawrence Livermore model (see Table~\ref{tab:eventsummary}) at distances of 10 kpc (5 kpc) would require a 0.45 (1.6) Mton background free detector to statistically compete with IceCube.
The second  GEANT GCALOR-based simulation combines all the steps in one program, which allows one e.g. to determine a 20\,\% dependence of the detector sensitivity on the incoming neutrino direction for neutrino interactions on electrons. Fig.~\ref{xyplot} shows the clustering of detected inverse beta neutrino interactions at the position of the detector strings to visualize the effective volumes.
\section{Performance Simulations}\label{sec:physicsperf}
All simulations are performed for the final IceCube array with 4800 standard and 360 high efficiency DOMs. We assume that 2\,\% of the DOMs are excluded from the analysis, either because they are not working or they give unstable rates.  
A likelihood ratio method was used to determine the range within which models can be distinguished. From sets of several thousand test experiments, we determine limits at the 90\,\% confidence level, while requiring that the tested scenario is detected in at least 50\,\% of the cases. Note that the ranges obtained should be interpreted as optimal as we assume that the model shapes are perfectly known and only the overall flux is left to vary; we also disregard the possibility that multiple effects, such as matter induced neutrino oscillations and neutrino self-interactions, could co-exist and thus may be hard to disentangle.  

\begin{figure*}[ht]
\centering
\includegraphics[angle=0,width=0.48\textwidth,bb=50 -50 600 380]{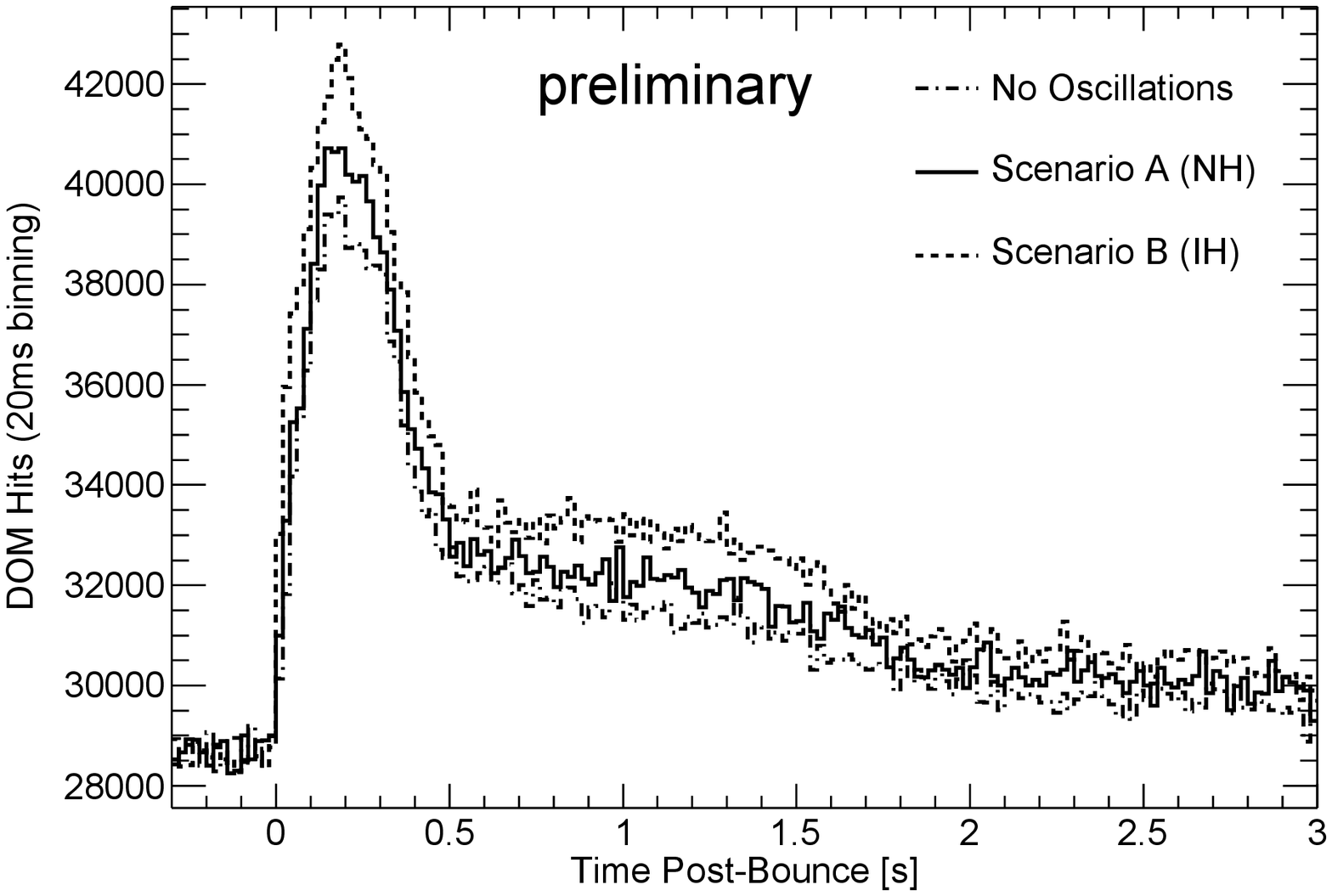}
\includegraphics[angle=0,width=0.48\textwidth,bb=-20 -50 530 380]{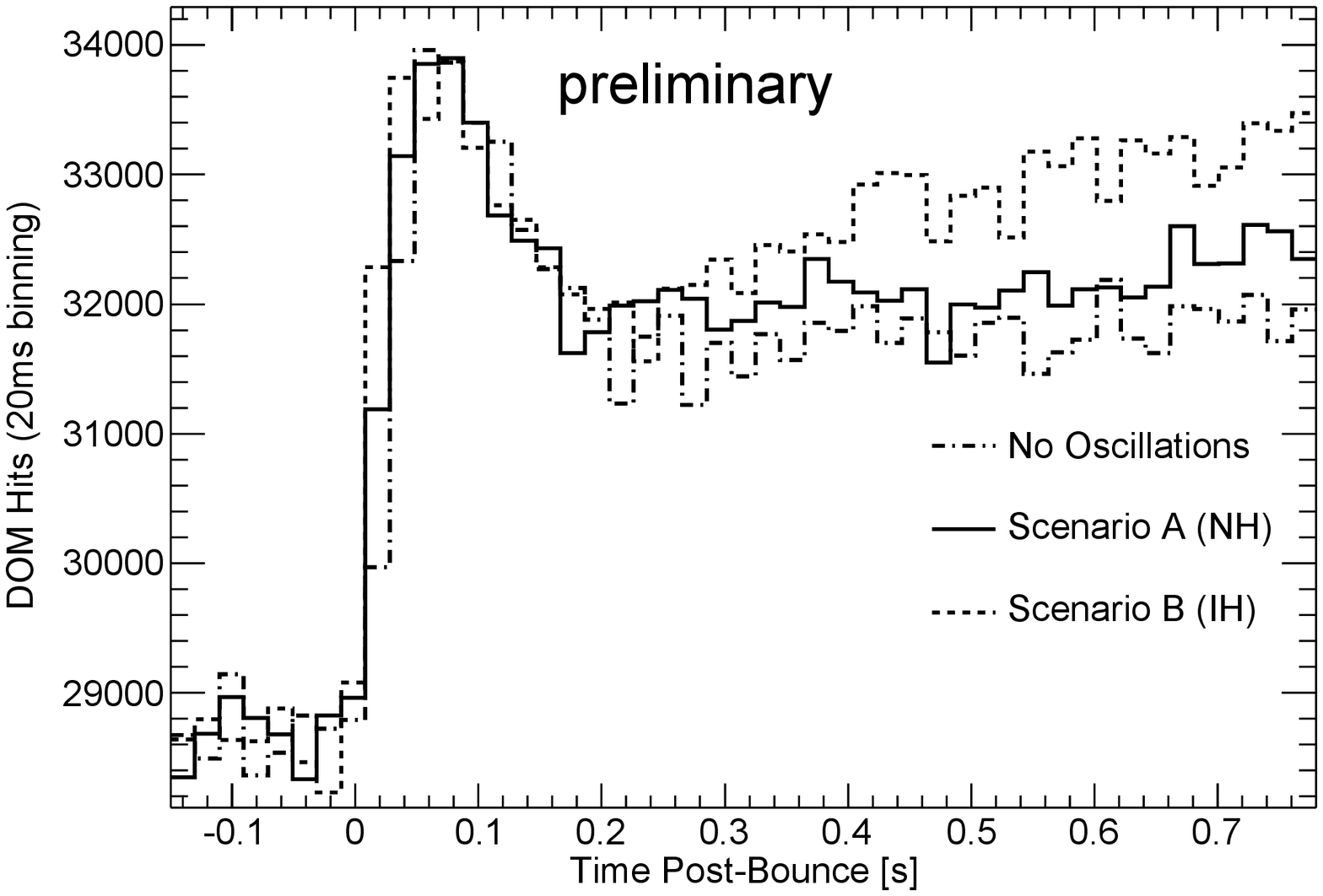}
\includegraphics[angle=0,width=0.45\textwidth, bb=50 -50 555 379]{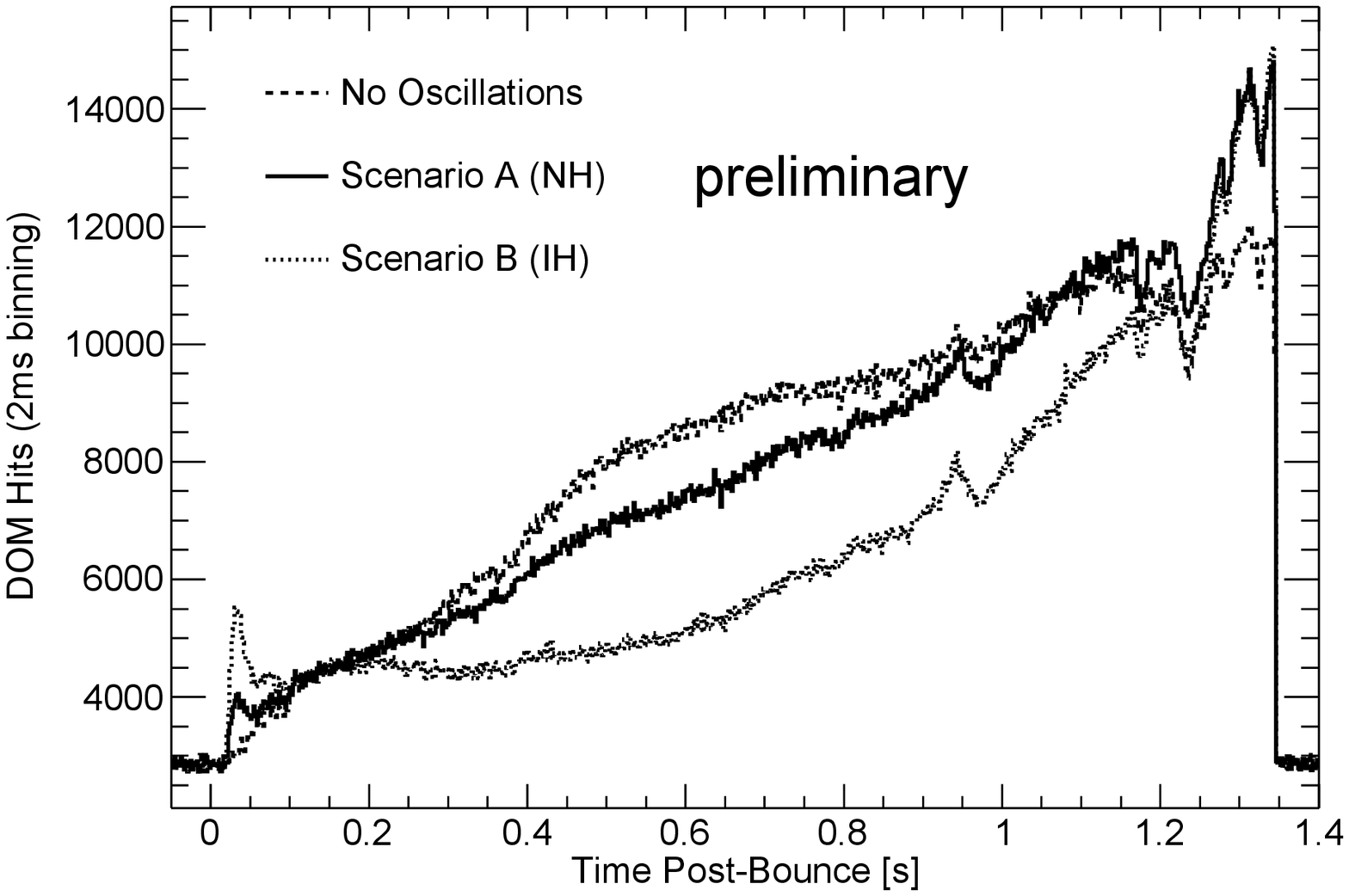}
\includegraphics[angle=0,width=0.51\textwidth,bb=-20 -50 558 378]{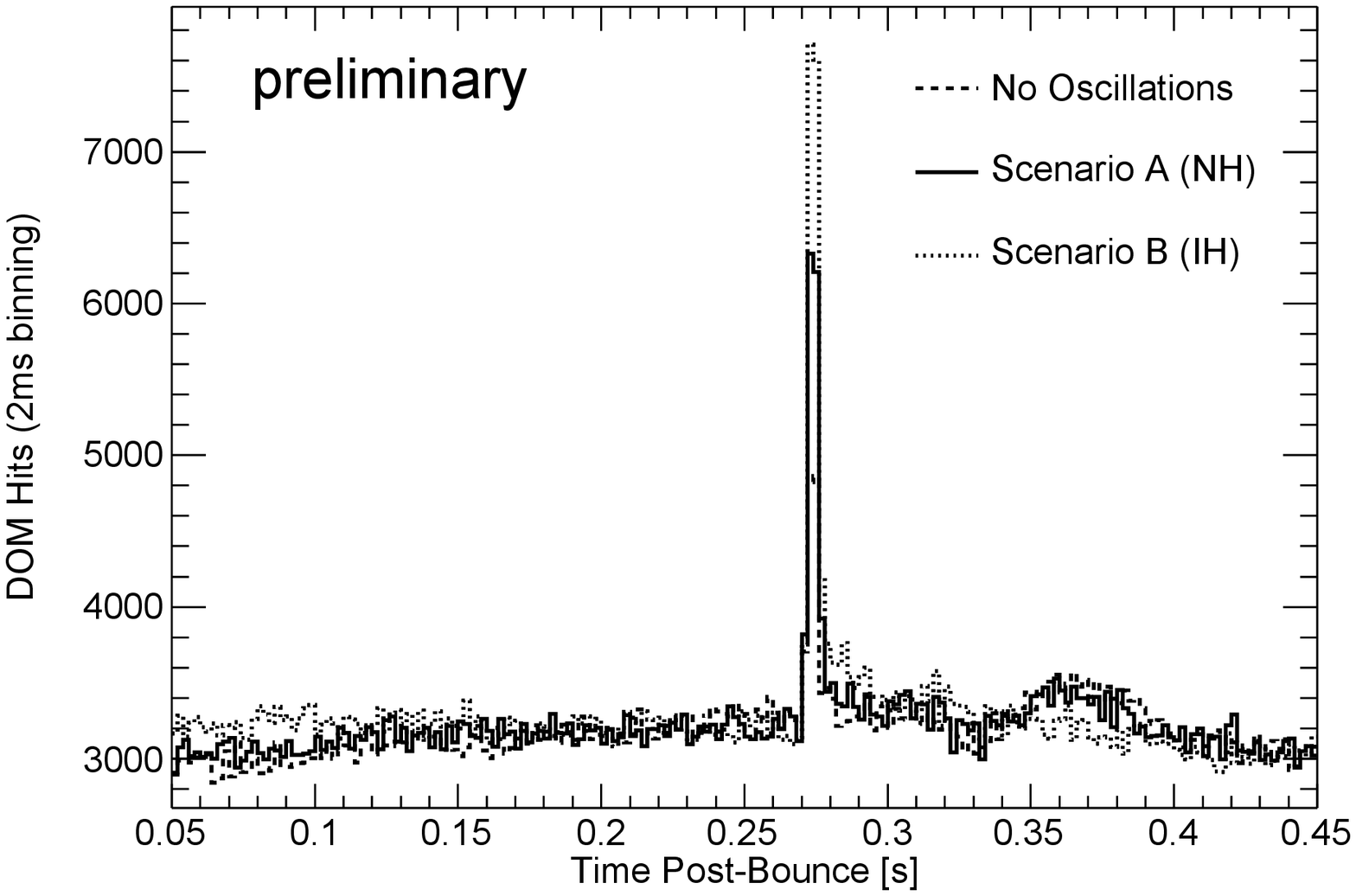}
\caption{Top: Expected rate distribution at \unit[10]{kpc} supernova distance for oscillation scenarios A (normal hierarchy) and B (inverted hierarchy). Fluxes and energies in the left plot are taken from the Lawrence-Livermore model~\cite{AstropPhys.496.216} and in the right plot from the Garching model~\cite{AstrAstrop.450.1}. The case of no oscillation is given as a reference.
Bottom:  Expected neutrino signal from the gravitational collapse of a non rotating massive star of 40 solar masses into a black hole following~\cite{Sumiyoshi}. Comparison of the neutrino light curve with quark-hadron phase transition for a progenitor star with 10 solar masses. The observation of the sharp $\bar\nu_\mathrm{e}$ induced burst 257 ms $< t< \, $261 ms after the onset of neutrino emission would constitute direct evidence of quark matter. The $1\,\sigma$-bands corresponding to measured detector noise (hatched area) have a width of about  $\pm$\, \unit[215]{counts} for a \unit[20]{ms} binning and $\pm$\,\unit[70]{counts} for a \unit[2]{ms} binning.}
\label{fig:rate}
\end{figure*}

IceCube is particularly well suited to study fine details of the neutrino flux as function of time. While the expected $\nu_\mathrm{e}$ deleptonization signal will be hard to disentangle with IceCube, forward and reverse shocks as well as stationary accretions shock instabilities (SASI) can be observed. 
Fig.~\ref{fig:rate} bottom left shows a simulation based on a model~\cite{Dasgupta-2010} that predicts a sudden spike in the $\bar\nu_\mathrm{e}$ flux lasting for a few ms while the neutron star turns to a quark star. The likelihood ratio test gives a deviation larger than 5~$\sigma$ from the hypothesis of no quark star formation for distances up to 30 kpc. The height and shape of the peak depend on the neutrino hierarchy. Scenarios A and B can be distinguished at 90\,\% C.L. up to distances of 30 kpc. 

\begin{figure}[ht]
\centering
\includegraphics[angle=0,width=0.54\textwidth]{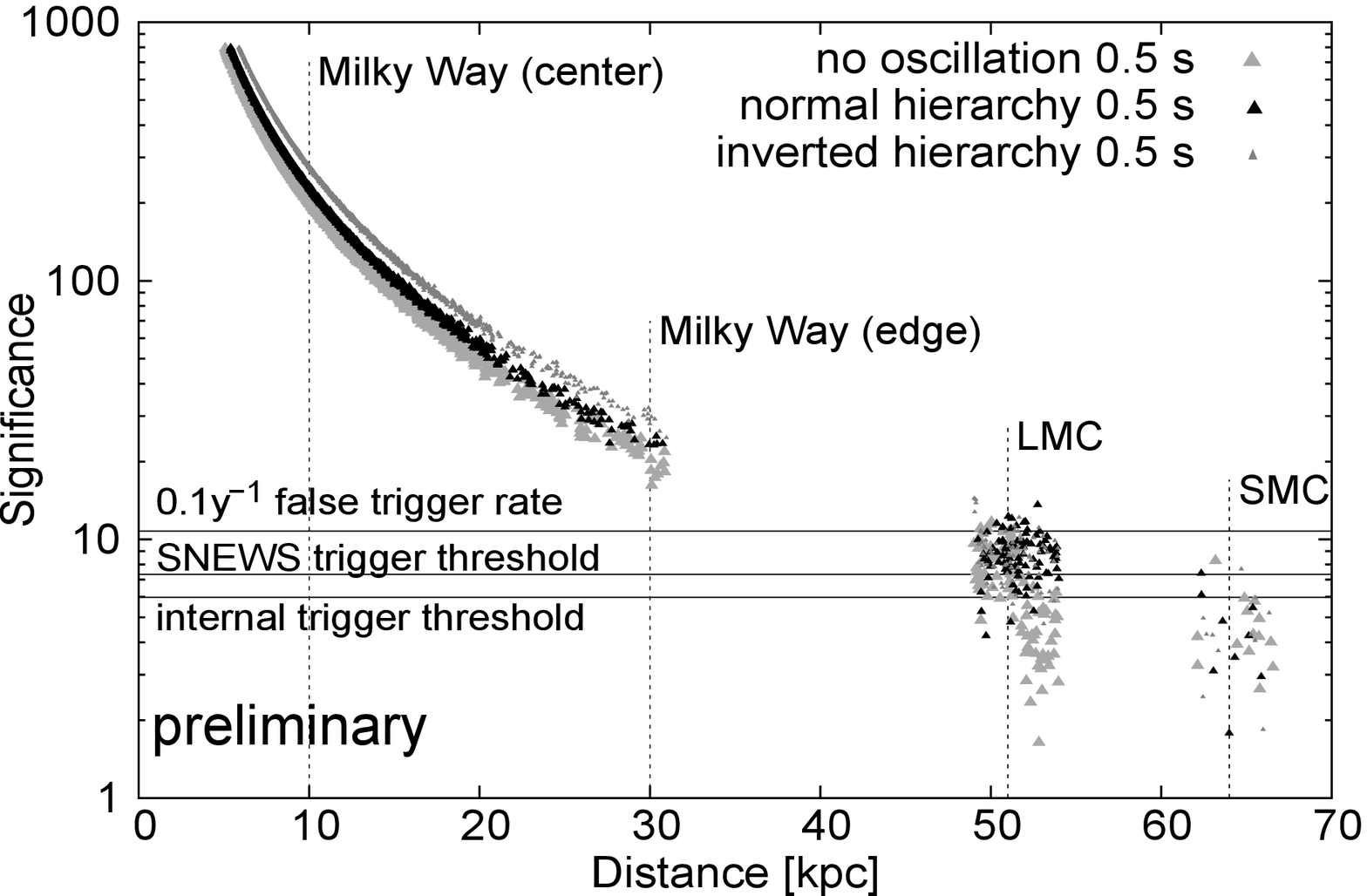}\quad\quad
\includegraphics[angle=0,width=0.40\textwidth,bb=0 20 564 551]{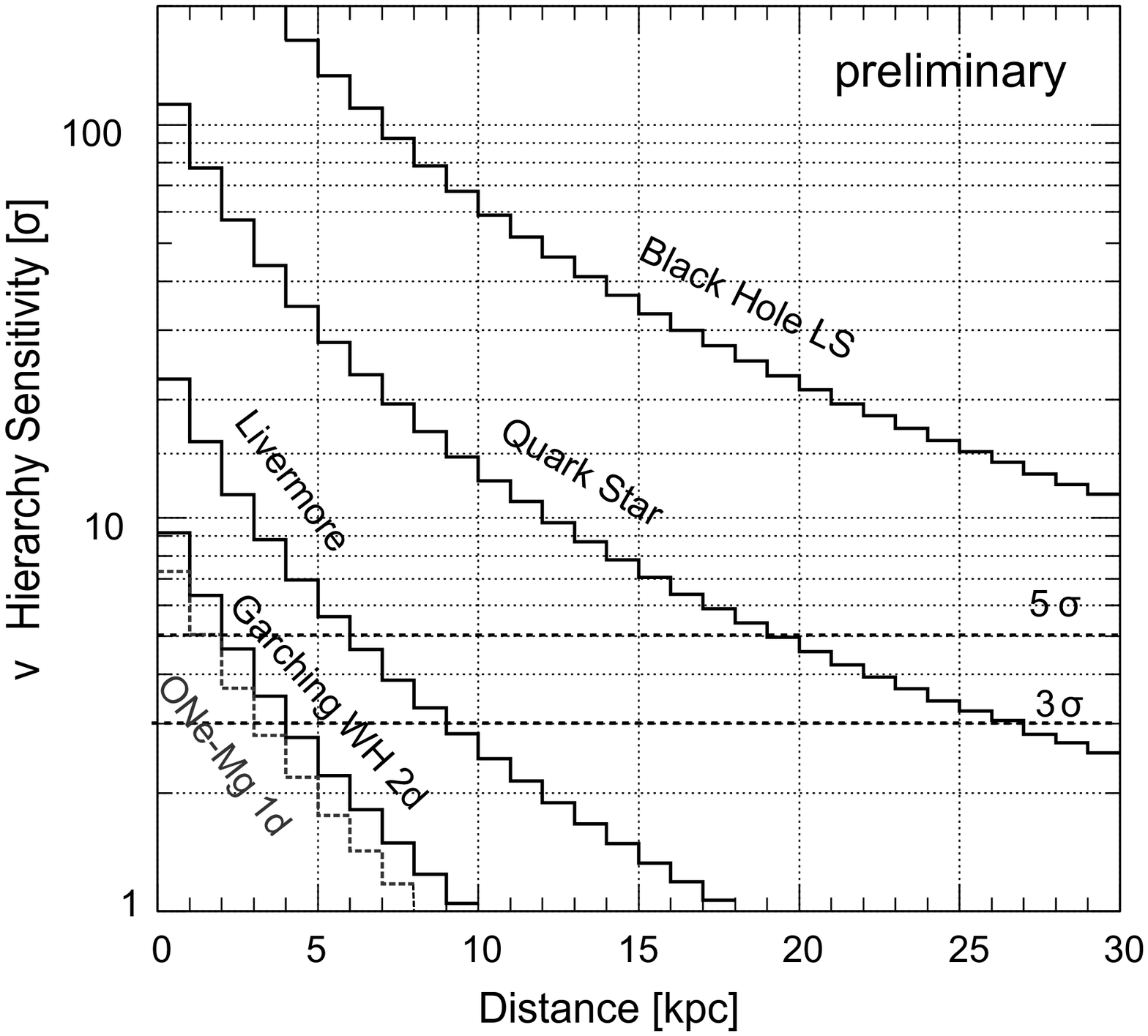}
\caption{Left: Significance versus distance 
assuming the Lawrence-Livermore model. The significances are increased by neutrino oscillations in the star by typically 15\,\% in case of a normal hierarchy (Scenario A) and 40\,\% in case of an inverted hierarchy (Scenario B). The Magellanic Clouds as well as center and edge of the Milky Way are marked. The density of the data points reflect the star distribution. Right: Number of standard deviation with which scenarios A (normal hierarchy) and B (inverted hierarchy) can be distinguished in at least 50\,\% of all cases as function of supernova distance for some of the models listed in Table~\ref{tab:eventsummary}. A likelihood ratio method was used assuming that the model shapes are perfectly known.}\label{fig:reach}
\end{figure}

Fig.~\ref{fig:rate} bottom right shows a simulation based on the prediction of~\cite{Sumiyoshi} for the formation of a black hole following a collapse of a 40 solar mass progenitor star. Neutrinos  reach energies up to 27 MeV ($\nu_\mathrm{e}$ and $\bar{\nu}_\mathrm{e}$) and 40 MeV ($\nu_{\mu}$ and $\nu_{\tau}$), carry a correspondingly large detection probability and thus produce very clear evidence for the formation of the black hole after 1.3 s.  The corresponding drop in $\nu$ rate can be identified at higher than 90\,\% C.L. for all stars in our Galaxy and the Magellanic Clouds. 
The simulation of an expected signal from a supernova within the Milky Way has to take into account the number of likely progenitor stars in the Galaxy as a function of the distance from Earth. The expected significances of supernova signals according to the Lawrence-Livermore model for three oscillation scenarios are shown in the left plot of Fig.~\ref{fig:reach}. 
%
The number of standard deviation with which normal and inverted $\nu$ hierarchies (Scenarios A and B) can be distinguished are plotted in Fig.~\ref{fig:reach} (right plot) as function of the supernova distance for selected models. The values represent the optimal cases when model shapes (but not necessarily the absolute fluxes) are perfectly known. 
Table~\ref{tab:eventsummary} lists the number of neutrino induced photon hits that would be recorded by IceCube on top of the nominal DOM noise level for various supernova models. 
\begin{table*}[ht]
\caption{\label{tab:eventsummary}Number of recorded DOM hits in IceCube ($\approx \# \nu$'s) for various models of the supernova collapse and progenitor masses assuming a distance of 10 kpc, approximately corresponding to the center of our Galaxy. A normal neutrino hierarchy is assumed.}
\centering
\vspace{2mm}
\begin{tabular}{llccc}
\hline\hline

Model & Reference & Progenitor       & $\#\nu$'s  & $\#\nu$'s   \\
      &           & mass ($M_\odot$) & $t<380$ ms & all times \\ 
\hline
``Livermore''                   & \cite{AstropPhys.496.216}						& 20        &$0.174\times 10^6$& $0.79\times 10^6$\\
``Garching LS-EOS 1d''          & \cite{AstrAstrop.450.1}   & $8-10$    &$0.069\times 10^6$ & -               \\
``Garching WH-EOS 1d''          & \cite{AstrAstrop.450.1}   & $8-10$    &$0.078\times 10^6$ & -               \\
``Garching SASI 2d''            & \cite{Marek}              & 15        &$0.106\times 10^6$ & -               \\
``1987A at 10 kpc''             & \cite{Pagliaroli-b}   & $15-20$   & &$(0.57\pm 0.18)\times 10^6$        \\
``O-Ne-Mg 1d''                  & \cite{Huedepohl}          & 8.8       &$0.054\times 10^6$ &$0.17\times 10^6$\\
``Quark Star (full opacities)'' & \cite{Dasgupta-2010}       &  10       &$0.067\times 10^6$ & -               \\
``Black Hole LS-EOS''           & \cite{Sumiyoshi}           &  40       &$0.395\times 10^6$ &$1.03\times 10^6$ \\
``Black Hole SH-EOS''           & \cite{Sumiyoshi}           &  40       &$0.335\times 10^6$ &$3.40\times 10^6$ \\
\hline
\end{tabular}
\end{table*}

\section{Summmary}
IceCube was completed in December 2010 and monitors $\approx$ 1 km$^3$ of deep Antarctic ice for particle induced photons with 5160 photomultiplier tubes. 
Since 2009 it supersedes AMANDA in the SNEWS network. With a 250 $\mu$s artificial dead time setting, the average DOM noise rate is 286 Hz. 
The data taking is very reliable and covers the whole calendar year, including periods when new strings were deployed. The uptime has continuously improved toward a goal of $>98$\,\% and reached 96.7\,\% in 2009. IceCube's sensitivity corresponds to a megaton scale detector for galactic supernovae, triggering on supernovae with about 200, 20, and 6 standard deviations at the galactic center (10 kpc), the galactic edge (30 kpc), and the Large Magellanic Cloud (50 kpc).  IceCube cannot determine the type, energy, and direction of individual neutrinos and the signal is extracted statistically from rates that include a noise pedestal. On the other hand, IceCube is currently the world's best detector for establishing subtle features in the temporal development of the neutrino flux. The statistical uncertainties at $10\un{kpc}$ distance in $20\un{ms}$ bins around the signal maximum are about 1.5\,\% and 3\,\% for the Lawrence Livermore and Garching models, respectively. 

Depending on the model, in particular the progenitor star mass, the assumed neutrino hierarchy and neutrino mixing, the total number of recorded neutrino induced photons from a burst 10 kpc away ranges between $\approx\, 0.17\times 10^6$  (8.8 M$_{\odot}$ O-Ne-Mg core), $\approx 0.8\, \times 10^6$ (20 M$_{\odot}$ iron core) to $\approx 3.4\, \times 10^6$ for a 40 M$_{\odot}$ progenitor turning into a black hole. 
For a supernova in the center of our Galaxy, IceCube's high statistics would allow for a clear distinction between the accretion and cooling phases, an estimation of the progenitor mass from the shape of the neutrino light curve, and for the observation of short term modulation due to turbulent phenomena or forward and reverse shocks during the cooling phase. 
IceCube will be able to distinguish inverted and normal hierarchies for the Garching, Lawrence-Livermore and black hole models for a large fraction of supernova bursts in our Galaxy provided that the model shapes are known and  $\theta_{13} > 0.9^\circ$. The slope of the rising neutrino flux following the collapse can be used to distinguish both hierarchies in a less model dependent way for distances up to 6 kpc at 90\,\% C.L. 
As in the case of the inverted hierarchy, coherent neutrino oscillation will enhance the detectable flux considerably. A strikingly sharp spike in the $\bar{\nu}_\mathrm{e}$ flux, detectable by IceCube for all stars within the Milky Way,  would provide a clear proof of the transition for neutron to a quark star as would be the sudden drop of the neutrino flux in case of a black hole formation.  

Further optimizations may be applied to the data acquisition and analysis in the future, e.g. by incorporating a more sophisticated method to remove correlated noise, by excluding the bin-by-bin contribution of measured cosmic ray muon hits to the rate measurement, by storing time stamps of all hits in case of a significant alarm to e.g. improve on the timing resolution, and by employing temporal templates in likelihood or cross-correlation studies.

\end{document}